\documentclass[conference,10pt]{IEEEtran}
\usepackage{float}
\usepackage{comment}
\usepackage[table]{xcolor}
\usepackage{cite}      
\usepackage{graphicx}  
\usepackage{psfrag}    
\usepackage{placeins}
\usepackage{lineno}
\usepackage{url}
\usepackage{stfloats}
\usepackage{subcaption}
\usepackage{caption}
\usepackage{times}
\usepackage{textcomp}
\usepackage{amsmath} 
\usepackage{amsfonts,amsthm,amssymb,mathtools}
\usepackage[font=small,labelfont=bf]{caption}
\usepackage{balance}
\usepackage[normalem]{ulem}
\usepackage{xcolor}

\usepackage{cuted}
\usepackage{amsmath,bm}
\usepackage{hyperref}
\usepackage{cleveref}

\begin{document}
\title{Radar Sensing using Dual-Beam Reconfigurable Intelligent Surface}
    
\author{\IEEEauthorblockN{Kainat~Yasmeen,  Shobha~Sundar~Ram, Debidas Kundu\\
Indraprastha Institute of Information Technology Delhi, New Delhi 110020 India\\
Email: \{kainaty, shobha, debidas\}@iiitd.ac.in}}

\maketitle

\begin{abstract}
Around-the-corner radar sensing offers an opportunity for the radar to exploit multipath scattering along walls to detect targets beyond blockages. However, the radar detection performance is limited to spotting uncooperative targets at specular angles. Recently, reconfigurable intelligent surfaces (RIS) involving metasurfaces with tunable unit cells have been researched for enhancing radar coverage around corners by directing beams towards non-specular angles. This article examines how practical considerations regarding the phase tuning of unit cells impact the RIS performance. 
Specifically, we examine the radar cross-section (RCS) obtained from two RIS configurations: In the first, each atom of the RIS is tuned based on a theoretical analog phase shift to realize idealized one-beam patterns at the desired angles. In the second configuration, each atom of the RIS is tuned based on a low-complexity, one-bit quantized element phase shift, which results in dual symmetric beams. The RIS configurations are then benchmarked with a metal plate of similar dimensions in both simulations and measurements. 
\end{abstract}

\providecommand{\keywords}[1]{\textbf{\textit{Keywords--}}#1}
\begin{IEEEkeywords}
Reconfigurable intelligent surfaces, Dual-beam RIS, radar cross-section

\end{IEEEkeywords}

\section{Introduction}
\label{sec:Introduction}
Radar technology plays a crucial role in detecting, tracking, and localizing targets in non-line-of-sight (NLOS) scenarios. For example, through-wall radar is used to detect and identify human activities behind walls \cite{ram2008doppler}, while ground-penetrating radar is used to detect buried objects \cite{cassidy2009ground}. These applications rely on electromagnetic signals at microwave frequencies, which penetrate the wall and ground materials. However, at higher frequencies, the dominant propagation phenomenon is the multipath scattering of signals along the surface of the walls, rather than through-wall propagation, due to the low skin depth of the building materials \cite{ram2010simulation,vishwakarma2020micro,10187182}. Hence, many recent works have focused on around-the-corner radars (ACR) that rely on single-bounce or multiple-bounce multipath signals along the walls to detect targets blocked from the radar's line of sight (LOS). This is useful for urban surveillance and for look-ahead sensing in advanced driver assistance systems, enabling cars to detect potential hazards beyond their LOS, while making turns \cite{rabaste2019detection}. However, the NLOS view of the radar in these cases is limited by the specular reflection of the signals by the walls, and the detection performance deteriorates significantly when the target is in non-specular regions. A potential solution that is being examined in these scenarios in recent literature is to use reconfigurable integrated surfaces (RIS) for detecting targets in NLOS regions at non-specular angles \cite{mercuri2023reconfigurable}.

An RIS is a flat two-dimensional metasurface array consisting of multiple unit cells that can be individually configured to control the phase, amplitude, frequency, and direction of the reflected wavefront by manipulating an incident electromagnetic wave. Since these reflective elements operate with low energy consumption, an RIS is a largely passive device \cite{basar2019wireless,10243495}.
Note that an RIS is distinct from a non-reconfigurable metasurface, where the meta-atoms are fabricated with fixed structural and geometric properties. As a result, these surfaces exhibit static interactions with incident electromagnetic waves, and their response cannot be modified post-fabrication \cite{aubry2021reconfigurable}. In contrast, RISs offer dynamic control over their interaction with radio waves. This adaptability is made possible by integrating electronic phase-changing materials — such as semiconductors or graphene —which serve as tunable reactive or resistive elements. These materials can be incorporated either between adjacent meta-atoms or within individual ones, acting as programmable switches \cite{liu2021reconfigurable,aubry2021reconfigurable}. 

There has been a tremendous focus in recent years on studying the applications of RIS for various purposes, such as wireless communications \cite{9206044,10715713}, wireless energy transfer \cite{zhao2020wireless}, and navigation \cite{wymeersch2020radio}. An RIS can be installed outside on building exteriors, interior ceilings, or walls. By dynamically configuring individual elements, an RIS can enhance spatial coverage and improve the signal quality in remote regions where the direct signal is not accessible \cite{liu2021reconfigurable}. Note that in all of these applications, the propagation between the transmitting source and receiver antennas is one-way forward scattering from the transmitter to the RIS to the receiver. There has been far less research focus on the application of RIS for radar \cite{liu2021reconfigurable,buzzi2021radar,buzzi2022foundations,mercuri2023reconfigurable}. In \cite{buzzi2021radar}, a theoretical study of the enhancement provided by an RIS for the radar detection of a single target was presented. This work was followed by a study in \cite{buzzi2022foundations} of the effectiveness of multiple-input multiple-output radar with RIS and the mitigation of multipath effects in indoor localization using RIS in \cite{mercuri2023reconfigurable}. Despite the promising benefits, practical constraints regarding the RIS beamwidth, bandwidth, and sidelobes have to be studied in detail before the widespread adoption of RIS for radar. One of the main practical challenges is incorporating phase shifts in the hardware of the unit cells in the RIS. In real-world systems, it is generally only feasible to incorporate quantized phase shifts from a discrete set of values. Hence, a low complexity RIS involves incorporating a one-bit phase quantization of the unit atoms of the metasurface. 
Furthermore, unlike point-to-point communications, radar applications involve detecting and tracking multiple uncooperative targets. Hence, RISs must be capable of directing multiple beams with low hardware and algorithm complexity. Finally, radar sensing involves two-way propagation, and hence the reciprocity of the RIS in the forward (radar to RIS to target) and reverse (target to RIS to radar) paths is an important aspect that needs to be studied. 

In this work, we propose using a dual-beam RIS for radar sensing of two targets by employing a beam steering method that utilizes low-complexity one-bit quantization of the phase shifts at the unit cells. We present both simulation and measurement data collected using a narrowband radar. Then, we study the accuracy of the RIS in directing beams towards different directions at various incident angles. We also compute the radar cross-sections (RCSs) of the RIS during forward and reverse scattering to test the reciprocity of the scattering phenomenology which is useful for determining the overall signal-to-noise ratio of the RIS-enhanced radar system. We benchmark the results of our dual-beam RIS with a metal plate and an ideal single-beam RIS with unquantized phase tuning. 

\indent The organization of this paper is as follows: the theory is presented in Section. \ref{sec:researchmethodology}. The simulation setup is discussed in Section.\ref{sec:simulationsetup}. The results are discussed in Section.\ref{sec:sim_results} and \ref{sec:mes_result} while the future scope of the work is discussed in the conclusion.
\section{Theory}
\label{sec:researchmethodology}
\begin{figure}[t]
\centering
\includegraphics[scale=0.8]{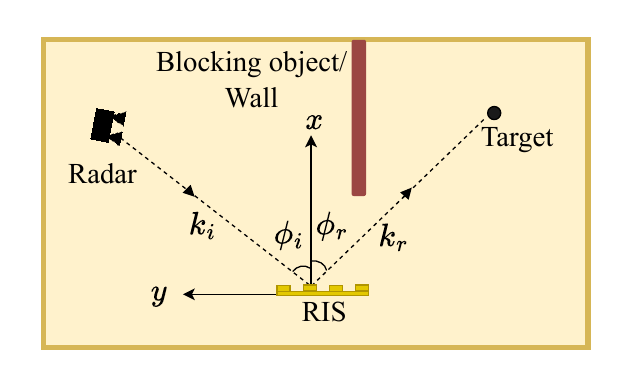}
\caption{Illustration of RIS-based Radar system for target detection. }
\label{fig:illustration}
\end{figure}

The objective of our work is to detect multiple targets using a dual-beam RIS configuration. For simplicity, we have considered a two-dimensional (2D) problem with a monostatic radar to detect two targets using RIS. We assume that the targets are blocked from the radar's LOS as shown in Figure.\ref{fig:illustration}.

We consider a one-dimensional (1D) RIS consisting of a linear array with N unit cells, spaced $\Delta y$ along the $y$ axis, as shown in the figure. 
We assume a transverse magnetic electric field, $\mathbf{E}_i$, incident upon the $n^{th}$ element of RIS at angle, $\phi_i$, as shown in
\par\noindent\small
\begin{align}
   \mathbf{E}_i = \hat{z}E_0e^{-j\mathbf{ k}_i \cdot \mathbf{r}_{n}}, \quad
\mathbf{k}_i = k_0 \begin{bmatrix}
        -\cos\phi_i & -\sin\phi_i & 0
    \end{bmatrix}
\end{align}
\normalsize
where $E_0$ is a constant and $\mathbf{r}_{n}$ is the position vector of $n^{th}$ element of RIS and $\mathbf{k}_i$ is the incident wave vector with $k_0$ propagation constant. The reflected electric field, $\mathbf{E}_r$, along angle, $\phi_r$, from each $n^{th}$ element is 
\par\noindent\small
\begin{align}
\mathbf{E}_r = \hat{z} \Gamma E_0 e^{-j \mathbf{k}_r \cdot \mathbf{r}_{n}} \quad
\mathbf{k}_r = k_0 \begin{bmatrix}
        \cos\phi_i & -\sin\phi_i & 0
    \end{bmatrix}.
\end{align}
\normalsize
where $\Gamma = -1$ is the reflection coefficient of the flat planar surface for perpendicular polarization and $\mathbf{k}_r$ is the reflected wave vector. The corresponding incident and reflected magnetic fields, for a wave impedance of $\eta_0$, are:
\par\noindent\small
\begin{align}
\mathbf{H}_i = \frac{E_0}{\eta_0}\left(-\hat{x}\sin\phi_i  + \hat{y}\cos\phi_i \right)e^{-j\mathbf{k}_i \mathbf{r}_{n}}
\\
\mathbf{H}_r = \frac{\Gamma E_0}{\eta_0}\left(-\hat{x}\sin\phi_i - \hat{y}\cos\phi_i\right) e^{-j \mathbf{k}_r \cdot \mathbf{r}_{n}}.
\end{align}
\normalsize
The surface current density on the RIS element (at $x=0$) is - 
\begin{align}
    \mathbf{J}_n = \hat{z} J_z = \hat{x} \times (\mathbf{H}_r + \mathbf{H}_i)
\end{align}
Using far-field array analysis \cite{balanis2024balanis}, we compute the far field scattered electric field from the entire RIS for any angle $\phi_s$ using
\begin{align}
\label{eq:Es}
\mathbf{E}_s(\phi_s) = \hat{z}\frac{jk_0E_0 \cos \phi_i}{\pi \sqrt{\rho}} \sum_{n=1}^{N} e^{-j(\mathbf{k}_s \cdot \mathbf{r}_n - \zeta_n)} \Delta y,
\end{align}
where $\mathbf{k}_s$ is the propagation vector along $\phi_s$ and $\rho$ is the distance of the center of RIS from the target. Note that due to the 2D problem statement, we consider cylindrical wave propagation rather than spherical waves. 
The tuning parameter for the RIS is the phase shift, $\zeta_n$, applied to each $n^{th}$ element across the RIS surface.

The generalized Snell's law can be expressed as \cite{sahoo20231}:
\begin{align}
    \sin \phi_r - \sin \phi_i = \frac{1}{k_0 \mu} \frac{\zeta_n}{(n-1)\Delta y},
    \label{eqn:one_beam}
\end{align}
where $\mu$ is the refractive index of the medium (which is assumed to be 1, corresponding to free space in our case).
In the case of a \emph{linear metal sheet}, $\Delta y = 0$, $\zeta_n = 0$ for infinite $N$, and the summation in \eqref{eq:Es} becomes an integral across the length of the plate. As a result, the mainlobe occurs at the specular angle where $\phi_r = \phi_i$.
In a \emph{single-beam RIS} scenario, $\zeta_n$ can be chosen such that the mainlobe occurs at a desired $\phi_r = \phi_d$. Theoretically, the mainlobe can be steered to any $\phi_d$ within the field of view with maximum gain, without the occurrence of grating lobes.
However, in real-world scenarios, $\zeta_n$ can only be realized in hardware for some specific discrete set of values, $\overline{\zeta}_n$. Hence, the ideal single-beam RIS is challenging to implement, and the error in the steering of the main lobe along the desired $\phi_d$ is a function of the quantization error - $||\zeta_n-\overline{\zeta}_n||^2$. 

For a \emph{dual-beam} RIS, the $\zeta_n$  is quantized with one bit to two values based on the following conditions \cite{sahoo20231}:
\begin{align}
   \overline{\zeta}_{n} =
\begin{cases}
0^\circ, & \zeta_{n} \notin [90^\circ, 270^\circ) \\
180^\circ, & \zeta_{n} \in [90^\circ, 270^\circ)
\end{cases} 
\label{eqn:dual_beam}
\end{align}
As a result of the 1-bit quantization, dual beams (mainlobe and grating lobe) will occur at $\phi_r = [+\phi_d,-\phi_d]$ at normal incidence.

Using \eqref{eq:Es}, we compute the radar cross section (RCS) of the RIS, which is the ratio of the scattered power along $\phi_d$ due to the incident power from $\phi_i$:
\begin{equation}
    \sigma(\phi_i,\phi_d) = 2\pi\rho\frac{|\mathbf{E}_s(\phi_d)|^2}{|\mathbf{E}_i(\phi_i)|^2}.
    \label{eqn:RCS}
\end{equation}
We use \ref{eqn:RCS} to calculate the RCS of the forward path, $\sigma_f(\phi_i,\phi_d)$, from the radar transmitter to the RIS to the target for incident angle, $\phi_i$, and scattering angle, $\phi_s$. Similarly, using the same RIS configuration (values of $\overline{\zeta}_n$), we also compute the electric field scattered back at the radar receiver and the corresponding RCS of the reverse path, $\sigma_r(\phi_d,\phi_i)$, from the target to the RIS to the monostatic radar receiver to evaluate the reciprocity of the RIS configuration for radar. 

Based on the two-way propagation of the signal along the forward and reverse paths, the received signal power at the radar for transmit power of $P_{tx}$, transmitting and receiving antenna gains of $G_{tx}$ and $G_{rx}$ and target RCS of $\sigma_t$ is given by 
\par\noindent\small
\begin{align}
\label{eq:radarrangeeqn}
P_{rx} = \frac{P_{tx}G_{tx}G_{rx}\sigma_f\sigma_r\sigma_t\lambda^2}{(4\pi)^5r_1^4r_2^4}.
\end{align}
\normalsize
Here, $r_1$ and $r_2$ are the distances of the RIS from the radar and target, respectively. While the path loss associated with this equation is naturally greater than that of free space LOS path between radar and target, the received power is boosted by the $\sigma_f$ and $\sigma_r$ when compared to multipath-based scattering from a wall in an ACR scenario. 
\section{Simulation Setup}
\label{sec:simulationsetup}
\begin{figure}[htbp]
\centering
\includegraphics[scale=0.67]{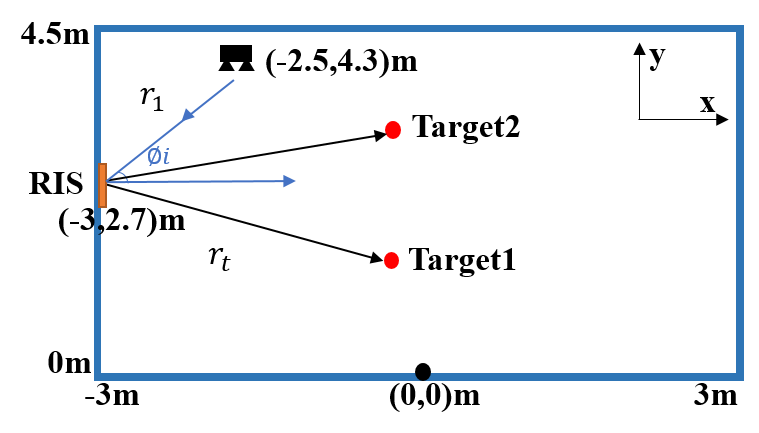}
\caption{Simulation setup for Radar, RIS, and multiple targets.}
\label{fig:simulation_setup}
\end{figure}

This section outlines the simulation setup for studying the RIS-based radar system. 
The simulation is carried out within a 2D environment on the $xy$ plane as shown in Fig.~\ref{fig:simulation_setup}. We consider a monostatic radar operating at 5.5 GHz located at $(-2.5, 4.3)$ m.
The RIS is oriented along the $y$-direction and consists of 16 elements spaced 0.016 m apart, centered at $(-3, 2.7)$ m. We  consider a two-target scenario in which each target performs a circular motion around a fixed center position described by:
\par\noindent\small
\begin{align}
x_i(t) &= x_{\text{center},i} + r \cos(\omega_i t), i = 1,2  \\
y_i(t) &= y_{\text{center},i} + r \sin(\omega_i t), i = 1,2.
\end{align}
\normalsize
Here, $(x_{\text{center},i}, y_{\text{center},i})$ is the center of the circular path, $\omega_i$ is the angular velocity, and $t$ is time. The $(x_{\text{center},1}, y_{\text{center},1}) = (-1,\, 2.1)$ m for the first target and $(-1,\, 3.2)$ m and $r = 0.2$ m over a duration of 1.5 seconds. The angular velocities are $\omega_1 = 2\pi,\text{rad/s}$ and $\omega_2 = 4\pi\,\text{rad/s}$ for both targets.

The received signal at the radar, $s(t)$, is simulated based on the theory described in the previous section and further processed to generate a joint time-frequency spectrogram.  This is obtained by applying the short-time Fourier transform (STFT) to $s(t)$ as shown in
\par\noindent\small
\begin{align}
  \textbf{STFT}(t,f) = \int_{} s(\tau)h(t-\tau)e^{-j2\pi f\tau}d\tau, 
\end{align}
\normalsize
where \( h(t) \) is the window function with a duration of 0.1 seconds \cite{ram2008simulation}. 

We consider three scenarios to evaluate the RIS performance. First, we consider a metal reflector of identical dimensions (length) to the RIS to observe passive reflection without control. In the second scenario, we consider an idealized single-beam RIS where the tuning parameters $\zeta_n, n=1\cdots N$ are not quantized. In the third scenario, we consider one-bit quantization of $\overline{\zeta}_n, n=1\cdots N$ as discussed previously to generate two simultaneous beams. 
\section{Simulation Results}
\label{sec:sim_results}
\begin{figure*}[t]
\includegraphics[width=500pt,height=270pt]{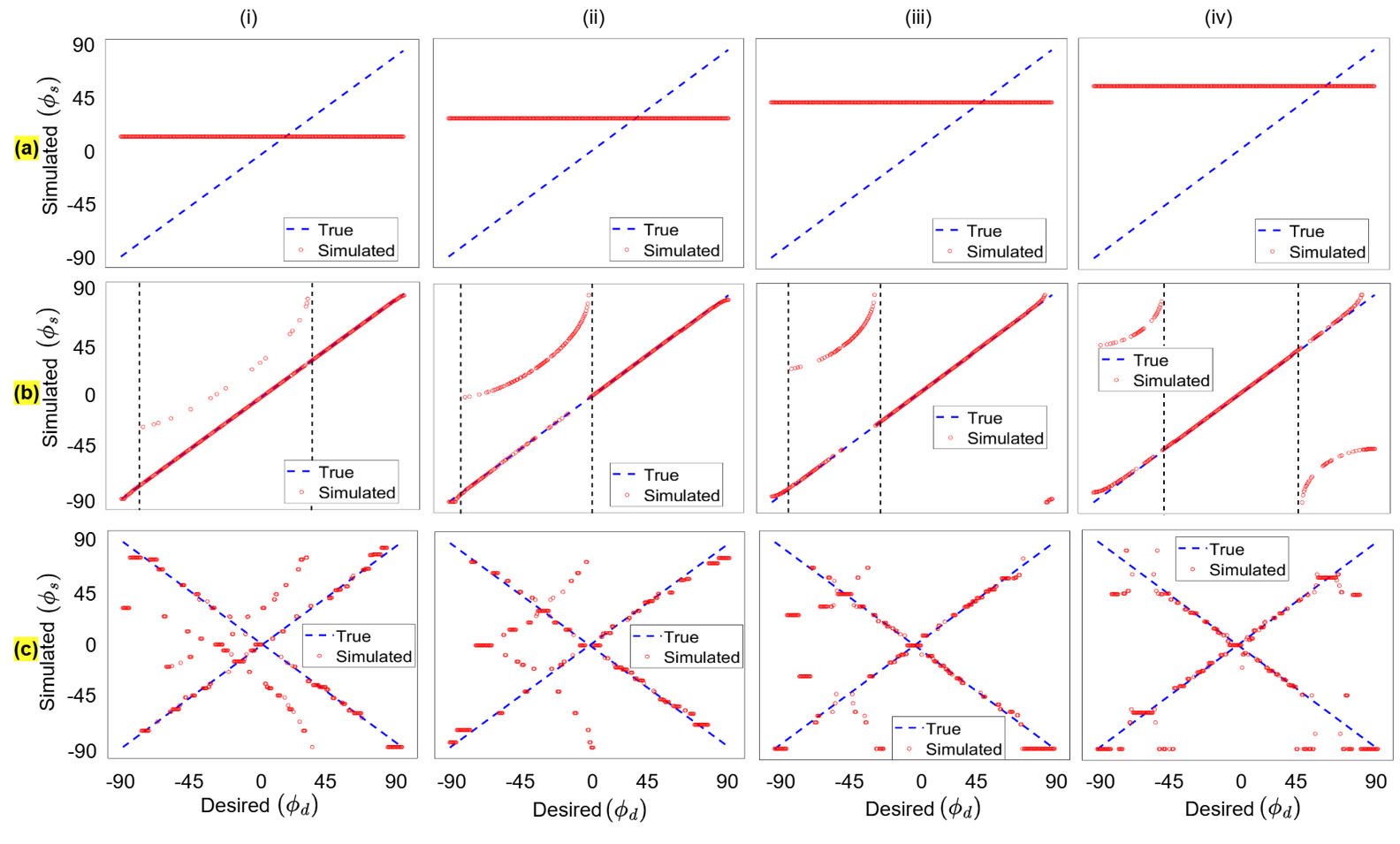}
\caption{(a) Metal, (b) One-beam, (c) Dual-beam RIS with incident angles of (i) $-15^\circ$, (ii) $-30^\circ$, (iii) $-45^\circ$, and (iv) $-60^\circ$ respectively.}
\label{fig:phis_vs_phid}
\end{figure*}

In this section, we validate the effectiveness of the proposed RIS-based radar by benchmarking its performance with that of a metal plate. First, we evaluate the effectiveness of the single-beam and dual-beam RISs in directing the main lobe towards the desired direction. The results are presented in Fig.\ref{fig:phis_vs_phid} along three rows.
We consider four distinct cases with  $\phi_i = -15^\circ$, $-30^\circ$, $-45^\circ$, and $-60^\circ$ along the four columns of the figure. For each $\phi_i$ case, we tune the single-beam and dual-beam RIS parameters to steer the beam from $-90^\circ$ to $90^\circ$. The $\phi_d$ is shown as a blue dotted line in the subfigures. For each $\phi_d$, we plot the position of the mainlobe in red. 

The top row shows the results for the metal surface. Here, even when $\phi_d$ is changed from $-90^\circ$ to $90^\circ$, the mainlobe of the RIS scattering is always directed towards the specular angle. Therefore when $\phi_i = 15^{\circ}$ the main lobe is also at $15^{\circ}$ in Fig.\ref{fig:phis_vs_phid}.(a,i). The same is true for other incident angles in columns (ii) to (iv) (Figs.\ref{fig:phis_vs_phid}a.ii-iv). The results show that the metal plate is unable to steer the reflected beam along non-specular angles. The middle row depicts the results of the single-beam RIS with idealized $\zeta_n$ without quantization. Here, we observe that the RIS can mostly steer the reflected beam toward the desired angle $\phi_d$ for all four $\phi_i$ since the red points overlap with the underlying dotted blue line. However, there are some regions of $\phi_d$ where the reflection does not show up. For example, for $\phi_i=45^{\circ}$, we see a gap near $\phi_d = -45^{\circ}$ in Fig.\ref{fig:phis_vs_phid}(b.ii). Also between $\phi_d = -90^{\circ}$ and $\phi_d = -45^{\circ}$, we observe grating lobes at other scattering angles. Similar trends are observed in the other figures in the row. In fact, in Fig.\ref{fig:phis_vs_phid}(b.iv) we see the emergence of two sets of grating lobes for greater values of $\phi_d$.  
The bottom row corresponds to the dual-beam configuration using one-bit quantized $\overline{\zeta}_n$ for RIS elements. This results in the generation of grating beams for every $\phi_d$. The dual-beam RIS also demonstrates overall greater error in steering the beam towards the desired direction. Similarly, we observe a notable increase in grating lobes across all four incident angles. 
\begin{figure}[htbp]
\centering
\includegraphics[width=250pt,height=112pt]{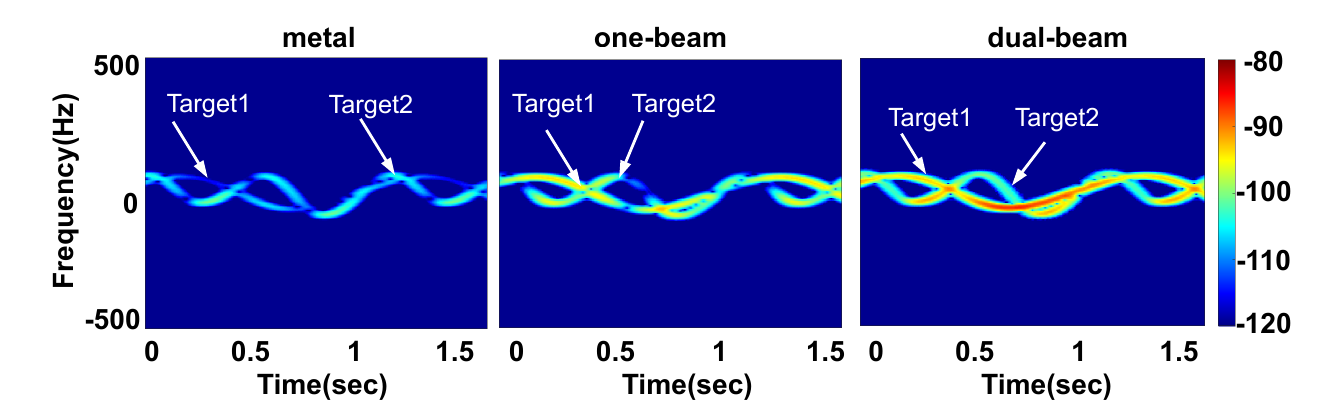}
\caption{Spectrograms of targets for (a) Metal, (b) One-beam RIS, and (c) dual-beam RIS .}
\label{fig:spectrogram}
\end{figure}

Next, we calculate the RCS for both the metal plate and the RIS for both the forward path, $\sigma_f$, and reverse path $\sigma_r$ in Table~\ref{table:simulated_RCS}.

\begin{table}[!htbp]
\centering
\renewcommand{\arraystretch}{1.3}
\caption{Simulated RCS values ($\sigma_f$, $\sigma_r$) in dBsm for metal, one-beam RIS, and dual-beam RIS under various incidence $(\phi_i)$ and desired $(\phi_d)$ angles}
\begin{tabular}{|p{0.55cm}|p{0.55cm}|cc|cc|cc|}
\hline
$(\phi_i)$&$(\phi_d)$ 
& \multicolumn{2}{c|}{Metal} 
& \multicolumn{2}{c|}{One-beam RIS} 
& \multicolumn{2}{c|}{Dual-beam RIS} \\
\hline
& & $\sigma_f$ & $\sigma_r$ 
  & $\sigma_f$ & $\sigma_r$ 
  & $\sigma_f$ & $\sigma_r$ \\
\hline
-$30^\circ$ &$30^\circ$ &-5.50 &-5.50  &-5.50  & -5.50 &-10.68  &-10.68 \\
$0^\circ$ & $30^\circ$   &-16.18 &-22.03 &1.92   &-5.50 &-1.92   &-17.62 \\
$0^\circ$ & $45^\circ$   &-19.31 &-22.89 &1.92   &-6.72 &-1.61   &-4.75 \\
\hline
\end{tabular}
\label{table:simulated_RCS}
\end{table}

\begin{table}[!htbp]
\centering
\renewcommand{\arraystretch}{1.3}
\caption{Experimental RCS values ($\sigma_f$, $\sigma_r$) in dBsm for metal and dual-beam RIS under various incidence $(\phi_i)$ and desired $(\phi_d)$ angles}
\begin{tabular}{|c|c|cc|cc|}
\hline
$(\phi_i)$ & $(\phi_d)$ 
& \multicolumn{2}{c|}{Metal} 
& \multicolumn{2}{c|}{Dual-beam RIS} \\
\hline
& & $\sigma_f$ & $\sigma_r$ 
  & $\sigma_f$ & $\sigma_r$ \\
\hline
$-30^\circ$ & $30^\circ$ & -7.20  & -7.20  & -12.20 & -12.20 \\
$0^\circ$ & $30^\circ$   & -23.20 & -17.20 & -3.20  & -10.20 \\
$0^\circ$ & $45^\circ$   & -21.40 & -16.20 & -13.20 & -10.20 \\
\hline
\end{tabular}
\label{table:experimental_RCS}
\end{table}

Our results show that when the desired angle, $\phi_d=30^{\circ}$, matches the incident angle in the top row, then $\sigma_f = \sigma_r$ for the metal plate, one-beam RIS, and dual-beam RIS. Furthermore, the RCS values of the one-beam RIS are the same as those of the metal plate. The dual-beam RIS has lower RCS (by 3 to 5 dB) than a one-beam RIS since the incident power is equally split between the two lobes. 
For non-specular reflection angles when $\phi_d \ne \phi_i$, then both $\sigma_f(\phi_i,\phi_d)$ and $\sigma_r(\phi_d,\phi_i)$ are significantly lower for the metal plate compared to the RIS. This is expected since the scattered field at $\phi_d$ arises due to the sidelobes rather than the main lobe, which is at the specular angle. In contrast, the RIS demonstrates much greater $\sigma_f$ and $\sigma_r$ for both one-beam RIS and dual-beam RIS, demonstrating the effectiveness of RIS for directing beams at non-specular angles. Interestingly, we also note that  $\sigma_f(\phi_i,\phi_d)$ and $\sigma_r(\phi_d,\phi_i)$ are not identical at these non-specular angles. indicating a deviation from strict reciprocity. This is attributed to the practical implementation of RIS phase profiles, which may not perfectly satisfy the conditions required for reciprocal scattering in non-specular directions.

Finally, we investigate the performance of RIS-assisted radar for detecting two rotating targets using different beam configurations and rotation speeds. In this scenario, both targets of similar cross-sections and ranges are located within an angular sector ranging from $-30^\circ$ to $+30^\circ$. The incident angle is fixed at $-45^\circ$. For the single-beam configuration, the RIS is steered to direct the reflected signal toward $30^\circ$. In contrast, the dual-beam configuration of the RIS simultaneously steers beams toward both $\pm30^\circ$, thereby covering a wider angular range. Figure~\ref {fig:spectrogram} shows that the micro-Doppler signatures of both targets are more clearly observed under the dual-beam setup in the third column. While the first target can be distinctly observed in both RIS configurations, the micro-Doppler corresponding to the second target appears significantly weaker in the single-beam case.

\section{Measurement Results}
\label{sec:mes_result} 
This section discusses the results obtained from measurement data collected with a one-bit dual-beam RIS and a metal plate of comparable dimensions. 
\subsection{Measurement Setup}
\begin{figure*}[!htbp]
\centering
\includegraphics[scale=0.6]{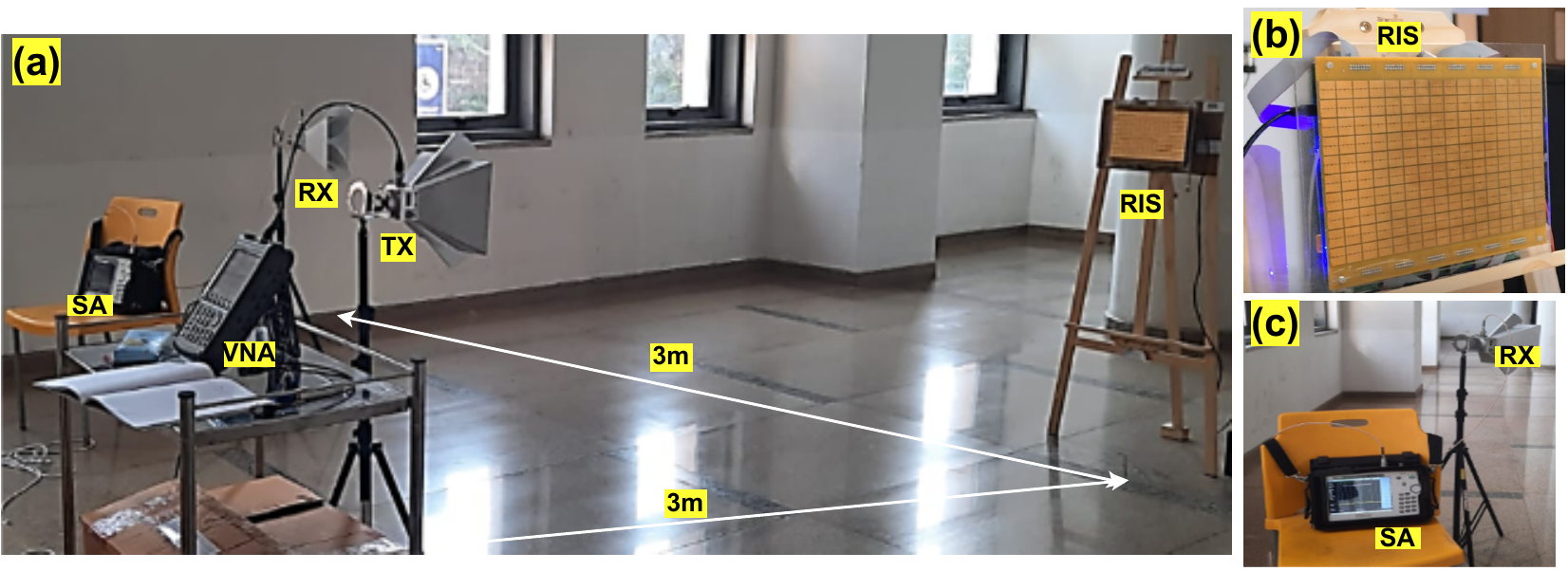}
\caption{(a) Experimental setup for beam-steering using RIS, (b) close-up view of RIS, and (c) close-up view of receiver setup.}
\label{fig:experimental_setup}
\end{figure*}

The measurement setup used to capture the forward scattered electric field in the desired $\phi_d$ for a given $\phi_i$ is shown in Fig.\ref{fig:experimental_setup}. A narrowband radar is configured using a FieldFox N9926A Vector Network Analyzer (VNA) operating at 5.5 GHz with a 100 kHz bandwidth, transmitting at a power level of +3 dBm through an HF907 horn antenna. The VNA and the RIS are positioned 3 m apart. A Spectrum Analyzer (SA), connected to a receiving horn antenna, is placed 3 m from the RIS on the opposite side to record the reflected signal, as shown in Fig.~\ref{fig:experimental_setup}(a) and (c).
\indent For this study, a custom-designed RIS of $256\times160$ mm with $16\times10$ unit cells was used as shown in Fig. \ref{fig:experimental_setup}(b). It uses a broadband 1-bit coding metasurface unit cell integrated with a PIN diode, offering a fractional bandwidth (FBW) of 18.51\% centered at 5.45 GHz, with a phase difference of $180^\circ \pm 20^\circ$ between its ON and OFF states. Note that this RIS supports only the dual-beam configuration, and hence we do not report measurement results for the one-beam configuration.
\subsection{Results}
We calculated the beam squint error for dual beam configuration for an incident angle $\phi_i=0^\circ$ and desired angle $\phi_d= \pm 45^\circ$. We observed the main lobe at an angle of $\phi_d'= \pm 46^\circ$.  The absolute error between the expected value and the measured value is given by $\left| \phi_d' - \phi_d \right| = \left| 46^\circ - 45^\circ \right| = 1^\circ.$

Next, we conduct experiments to calculate the $\sigma_f$ and $\sigma_r$. These are computed using  \eqref{eq:radarrangeeqn} modified for one-way propagation. The results are presented in Table \ref {table:experimental_RCS}. 
Similar to the simulation results, we again see that $\sigma_f = \sigma_r$ at specular angles for both the metal plate and the RIS. Again, the RCS of the dual-beam RIS is lower than that of the metal plate at these angles for two reasons. The first reason is that the radiated power is split between two main lobe beams, resulting in a reduced RCS along each beam. Second, the real RIS is characterized by additional ohmic losses that are not modeled in the simulations, as well as system losses associated with the use of cables and adapters. This results in an overall reduction of 2 dB in the RCS of the experimental results compared to the simulation results. 
Again, at non-specular angles, we observe that $\sigma_f$ and $\sigma_r$ of the RIS are greater than those of the metal plate due to steering of the main beam along the desired direction by the RIS, unlike the metal plate that radiates lower sidelobes along these angles. Also, consistent with the trend observed in the simulations, the measurement results show that exact reciprocity in $\sigma_f$ and $\sigma_r$ is not maintained at non-specular angles. 
\section{Conclusion}
\label{sec:conclusion}
This article explores one-beam and dual-beam RIS-based radar sensing. Unlike a metal surface that reflects main lobes at specular reflection angles, RIS configurations enable directional and multi-directional beamforming through the phase tuning of individual atoms in the metasurface. The one-beam pattern is realized by using exact phase shifts tuned based on generalized Snell's law to direct beams at desired directions. The dual-beam pattern utilizes a low-complexity one-bit quantized phased tuning. The advantage of this mechanism is that it is easy to build, resulting in two beams that can be used to detect two targets. The disadvantage is that these beams are symmetrical about the normal of the RIS with slightly lower RCS values. At specular reflection angles, the RIS looks identical to a metal plate. However, at desired non-specular angles, RIS offers higher RCS than the plate, and the forward and reverse RCS are not similar. The higher RCS of the RIS when compared to plates makes it potentially useful for ACR radar sensing.  

Numerical simulations and experimental results demonstrate that RIS-based approaches significantly enhance focusing and detection performance, with the dual-beam RIS offering improved coverage for multi-target scenarios. Future work will focus on experimental validation of target detection using both single- and dual-beam RIS, as well as minimizing the grating lobes of 1-bit coding metasurfaces using the prephasing technique.
\bibliography{Bibliography}

@article{basar2019wireless,
  title={Wireless communications through reconfigurable intelligent surfaces},
  author={Basar, Ertugrul and Di Renzo, Marco and De Rosny, Julien and Debbah, Merouane and Alouini, Mohamed-Slim and Zhang, Rui},
  journal={IEEE access},
  volume={7},
  pages={116753--116773},
  year={2019},
  publisher={IEEE}
}

@article{rabaste2019detection,
  title={Detection--localization algorithms in the around-the-corner radar problem},
  author={Rabaste, Olivier and Bosse, Jonathan and Poullin, Dominique and S{\'a}enz, Israel D Hinostroza and Letertre, Thierry and Chonavel, Thierry and others},
  journal={IEEE Transactions on Aerospace and Electronic Systems},
  volume={55},
  number={6},
  pages={2658--2673},
  year={2019},
  publisher={IEEE}
}

@article{ram2008doppler,
  title={Doppler-based detection and tracking of humans in indoor environments},
  author={Ram, Shobha Sundar and Li, Yang and Lin, Adrian and Ling, Hao},
  journal={Journal of the Franklin Institute},
  volume={345},
  number={6},
  pages={679--699},
  year={2008},
  publisher={Pergamon}
}

@inproceedings{ram2008simulation,
  title={Simulation of human microDopplers using computer animation data},
  author={Ram, Shobha Sundar and Ling, Hao},
  booktitle={2008 IEEE Radar Conference},
  pages={1--6},
  year={2008},
  organization={IEEE}
}

@inproceedings{vishwakarma2020micro,
  title={Micro-Doppler signatures of dynamic humans from around the corner radar},
  author={Vishwakarma, Shelly and Rafiq, Aaquib and Ram, Shobha Sundar},
  booktitle={2020 IEEE International Radar Conference (RADAR)},
  pages={169--174},
  year={2020},
  organization={IEEE}
}

@article{buzzi2021radar,
  title={Radar target detection aided by reconfigurable intelligent surfaces},
  author={Buzzi, Stefano and Grossi, Emanuele and Lops, Marco and Venturino, Luca},
  journal={IEEE Signal Processing Letters},
  volume={28},
  pages={1315--1319},
  year={2021},
  publisher={IEEE}
}

@article{buzzi2022foundations,
  title={Foundations of MIMO radar detection aided by reconfigurable intelligent surfaces},
  author={Buzzi, Stefano and Grossi, Emanuele and Lops, Marco and Venturino, Luca},
  journal={IEEE Transactions on Signal Processing},
  volume={70},
  pages={1749--1763},
  year={2022},
  publisher={IEEE}
}

@article{mercuri2023reconfigurable,
  title={Reconfigurable intelligent surface-aided indoor radar monitoring: A feasibility study},
  author={Mercuri, Marco and Arnieri, Emilio and De Marco, Raffaele and Veltri, Pierangelo and Crupi, Felice and Boccia, Luigi},
  journal={IEEE Journal of Electromagnetics, RF and Microwaves in Medicine and Biology},
  year={2023},
  publisher={IEEE}
}

@article{aubry2021reconfigurable,
  title={Reconfigurable intelligent surfaces for N-LOS radar surveillance},
  author={Aubry, Augusto and De Maio, Antonio and Rosamilia, Massimo},
  journal={IEEE Transactions on Vehicular Technology},
  volume={70},
  number={10},
  pages={10735--10749},
  year={2021},
  publisher={IEEE}
}

@article{liu2021reconfigurable,
  title={Reconfigurable intelligent surfaces: Principles and opportunities},
  author={Liu, Yuanwei and Liu, Xiao and Mu, Xidong and Hou, Tianwei and Xu, Jiaqi and Di Renzo, Marco and Al-Dhahir, Naofal},
  journal={IEEE communications surveys \& tutorials},
  volume={23},
  number={3},
  pages={1546--1577},
  year={2021},
  publisher={IEEE}
}

@article{zhao2020wireless,
  title={Wireless power transfer empowered by reconfigurable intelligent surfaces},
  author={Zhao, Long and Wang, Zhouyin and Wang, Xiaodong},
  journal={IEEE Systems Journal},
  volume={15},
  number={2},
  pages={2121--2124},
  year={2020},
  publisher={IEEE}
}

@article{wymeersch2020radio,
  title={Radio localization and mapping with reconfigurable intelligent surfaces: Challenges, opportunities, and research directions},
  author={Wymeersch, Henk and He, Jiguang and Denis, Benoit and Clemente, Antonio and Juntti, Markku},
  journal={IEEE Vehicular Technology Magazine},
  volume={15},
  number={4},
  pages={52--61},
  year={2020},
  publisher={IEEE}
}

@article{cassidy2009ground,
  title={Ground penetrating radar data processing, modelling and analysis},
  author={Cassidy, Nigel J and Jol, HM},
  journal={Ground penetrating radar: theory and applications},
  pages={141--176},
  year={2009},
  publisher={Elsevier Amsterdam}
}

@article{ram2010simulation,
  title={Simulation and analysis of human micro-Dopplers in through-wall environments},
  author={Ram, Shobha Sundar and Christianson, Craig and Kim, Youngwook and Ling, Hao},
  journal={IEEE Transactions on Geoscience and remote sensing},
  volume={48},
  number={4},
  pages={2015--2023},
  year={2010},
  publisher={IEEE}
}

@ARTICLE{10187182,
  author={Yasmeen, Kainat and Ram, Shobha Sundar},
  journal={IEEE Journal of Selected Topics in Applied Earth Observations and Remote Sensing}, 
  title={Estimation of Electrical Characteristics of Inhomogeneous Walls Using Generative Adversarial Networks}, 
  year={2023},
  volume={16},
  number={},
  pages={7009-7023},
  doi={10.1109/JSTARS.2023.3296872}}

@book{balanis2024balanis,
  title={Balanis' Advanced Engineering Electromagnetics},
  author={Balanis, Constantine A},
  year={2024},
  publisher={John Wiley \& Sons}
}

@inproceedings{sahoo20231,
  title={A 1-Bit Coding Reconfigurable Metasurface Reflector for Millimeter Wave Communications in E-Band},
  author={Sahoo, Deepak Kumar and Chakraborty, Chanchal and Kundu, Debidas and Patnaik, Amalendu and Chakraborty, Ajay and others},
  booktitle={2023 IEEE Microwaves, Antennas, and Propagation Conference (MAPCON)},
  pages={1--4},
  year={2023},
  organization={IEEE}
}

@ARTICLE{9206044,
  author={Tang, Wankai and Chen, Ming Zheng and Chen, Xiangyu and Dai, Jun Yan and Han, Yu and Di Renzo, Marco and Zeng, Yong and Jin, Shi and Cheng, Qiang and Cui, Tie Jun},
  journal={IEEE Transactions on Wireless Communications}, 
  title={Wireless Communications With Reconfigurable Intelligent Surface: Path Loss Modeling and Experimental Measurement}, 
  year={2021},
  volume={20},
  number={1},
  pages={421-439},
  doi={10.1109/TWC.2020.3024887}}

@ARTICLE{10715713,
  author={Deng, Min and Ahmed, Manzoor and Wahid, Abdul and Soofi, Aized Amin and Khan, Wali Ullah and Xu, Fang and Asif, Muhammad and Han, Zhu},
  journal={IEEE Transactions on Intelligent Vehicles}, 
  title={Reconfigurable Intelligent Surfaces Enabled Vehicular Communications: A Comprehensive Survey of Recent Advances and Future Challenges}, 
  year={2024},
  volume={},
  number={},
  pages={1-28},
  doi={10.1109/TIV.2024.3476934}}

@ARTICLE{10243495,
  author={Chepuri, Sundeep Prabhakar and Shlezinger, Nir and Liu, Fan and Alexandropoulos, George C. and Buzzi, Stefano and Eldar, Yonina C.},
  journal={IEEE Signal Processing Magazine}, 
  title={Integrated Sensing and Communications With Reconfigurable Intelligent Surfaces: From signal modeling to processing}, 
  year={2023},
  volume={40},
  number={6},
  pages={41-62},
  doi={10.1109/MSP.2023.3279986}}
\bibliographystyle{ieeetr}
\end{document}